\documentclass[12pt]{article}
\usepackage[T1]{fontenc}
\usepackage[latin1]{inputenc}
\usepackage{a4wide}
\usepackage{graphics}
\usepackage{psfrag}
\usepackage[dvips]{graphicx}
\usepackage{amssymb}
\usepackage{subfigure}
\makeatletter

\newcommand{\eqalign}[1]{
\null \,\vcenter {\openup \jot \ialign {\strut \hfil $\displaystyle {
##}$&$\displaystyle {{}##}$\hfil \crcr #1\crcr }}\,}
\newcommand{\be}{\begin{equation}}
\newcommand{\ee}{\end{equation}}
\newcommand{\ba}{\begin{array}}
\newcommand{\ea}{\end{array}}
\newcommand{\bea}{\begin{eqnarray}}
\newcommand{\eea}{\end{eqnarray}}

\newcommand{\bigx}[1]{\bBigg@{#1}}

\makeatother

\begin{document}
        
\title{{\normalsize \begin{flushright}\normalsize{ITP--Budapest Report 578}\end{flushright}\vspace{1cm}}
Some semi-classical issues in boundary sine-Gordon model}

\author{M. Kormos and L. Palla}

\maketitle
{\centering \emph{Institute for Theoretical Physics }\\
\emph{Roland Eötvös University, }\\
\emph{H-1117 Budapest, Pázmány sétány 1/A, Hungary}\par}

\begin{abstract}
The semi-classical quantisation of the two lowest energy static 
solutions of boundary sine-Gordon model is considered. A relation
between the Lagrangian and bootstrap parameters is established by
comparing their quantum corrected energy difference and the exact
one. This relation is also confirmed by studying the 
semi-classical limit of soliton reflections on the boundary. 
\end{abstract}
{\par\centering PACS codes: 64.60.Fr, 11.10.Kk   
\par}
\newpage

\makeatletter

\renewcommand\theequation{\hbox{\normalsize\arabic{section}.\arabic{equation}}} 

\@addtoreset{equation}{section} 

\renewcommand\thefigure{\hbox{\normalsize\arabic{section}.\arabic{figure}}} 

\@addtoreset{figure}{section} 

\renewcommand\thetable{\hbox{\normalsize\arabic{section}.\arabic{table}}} \@addtoreset{table}{section}

\makeatother

\section{Introduction}
The sine-Gordon model is one of the most extensively studied quantum
field theories. The interest stems partly from the wide range of
applications that extend from particle physics to condensed matter
systems and partly from the fact that many of the interesting physical
quantities can be computed exactly due to its integrability. All these
properties are inherited by the boundary sine-Gordon model (BSG)
obtained by restricting the ordinary one to the negative half line by
imposing appropriate, integrability preserving, boundary conditions at $x=0$ \cite{SK},
\cite{GZ}. 

The novel feature of BSG is the complicated spectrum of boundary bound
states manifesting themselves as appropriate poles in the various
reflection amplitudes \cite{GZ}-\cite{genpap}. These exact amplitudes are
obtained from solving the boundary versions of the Yang-Baxter,
unitarity and crossing equations \cite{GZ} in the bootstrap program
\cite{gosh}, \cite{FK}, \cite{patr}, \cite{genpap}. Therefore in the 
general case the reflection
factors and the spectrum of bound states depend on two \lq bootstrap'
or \lq infrared' parameters that characterize the solutions of these
equations. These parameters should be determined somehow by the two \lq
ultraviolet' or \lq Lagrangian' boundary parameters appearing in the
boundary potential enforcing the boundary condition. This question
leads then to the problem of establishing a relation between the exact
algebraic solution of the quantum theory and the classical
Lagrangian. A semi-classical quantisation of the classical theory may
provide the necessary link.

The quest for the relation connecting the two sets of parameters 
(also called UV-IR relation below) has a
long history. For Dirichlet boundary conditions, when only one
bootstrap and one Lagrangian parameters survive, it was obtained
already in \cite{GZ}. A general expression was given by
Al.B. Zamolodchikov \cite{Zupb} obtained from describing the BSG model
as a bulk and boundary perturbed conformal field theory, but
unfortunately these results remained unpublished. 
Recently 
some arguments were presented for
the general form of the UV
-- IR relation in \cite{genpap} by comparing the parameter
dependencies of some patterns (such as
global symmetries and ground state sequences) in the bootstrap
solution and in the classical theory. While this general form is
consistent with Zamolodchikov's solution, it leaves the coupling
constant dependency of a crucial coefficient undetermined. A TCSA
study of the spectrum of BSG in finite volume \cite{uvir} confirmed
that
Zamolodchikov's constant has the correct $\beta $ dependency. 
In contrast in the
boundary sinh-Gordon model the UV -- IR relation was determined by
Corrigan and Taormina by 
comparing the WKB and bootstrap spectra of breathers \cite{CT}. It
turns out after 
analytically continuing this relation to the sine-Gordon model, that
its general form is the expected one, but its coefficient 
depends on $\beta$ in a different way. 

Motivated by the above we consider  in this paper two problems in boundary
sine-Gordon model, where the semi-classical approximation can be
determined starting from the classical Lagrangian, and the results can
be compared to the appropriate limits of the exact solution. We choose
these problems to involve in one way or other the solitons in BSG, as
they have no analogues in sinh-Gordon theory, thus the results cannot
be obtained or predicted by a simple analytic continuation. 

The first problem we investigate is the 
semi-classically corrected  energy difference of the two lowest energy 
static solutions in boundary sine-Gordon model. These classical
solutions are in fact given by a static bulk soliton/antisoliton
\lq standing at the right place' \cite{SSW} \cite{genpap}, thus their semi-classical
quantisation amounts to the adaptation of the soliton quantisation
\cite{Raj} to the boundary problem. On the other hand these solutions
may be thought of as the classical analogues of the exact ground state
$\vert\rangle$, and the first excited boundary state $\vert 0\rangle$
respectively \cite{genpap}, thus the semi-classically corrected energy
difference should be compared to the limit of these two exact
energies. This leads then to a relation between the Lagrangian and the
bootstrap parameters.

The second problem we investigate is the semi-classical soliton
reflection on the boundary at $x=0$. The idea to compare the
semi-classical phase shift of this process - obtained from the
classical time delay - and the limit of the exact amplitude coming
from the algebraic solution was
suggested by Saleur, Skorik and Warner \cite{SSW}. Although they
determined the classical time delay in the general case (for ground
state boundary at least), they made the comparison for Dirichlet
boundary conditions only. Here we show that the comparison in the
general case leads to the same UV-IR relation we obtained from the
first problem.

The paper is organized as follows: the semi-classical quantisation of
the static solutions is carried out in sect. 2. The results are
compared to the limit of the exact solution in section 3. Section 4
is reserved for the investigation of the soliton reflection and we make
our conclusions in sect. 5.

\section{Semi-Classical quantisation of the static solutions}

In this section we carry out the semi-classical quantisation of two
static solutions in 
boundary sine-Gordon model and 
compute the semi-classical quantum correction to the
 difference between their classical energies. We
start by summarizing some known facts about this theory
and the classical solutions in question.

The boundary version of sine-Gordon model
is defined by the action  \cite{GZ}:
\begin{equation}
S=\int _{-\infty }^{\infty }dt\int _{-\infty }^{0}dx{\cal {L}}_{\rm
  SG}-\int _{-\infty }^{\infty }dtV_{\rm  B}(\Phi _{B}),\qquad \quad 
{\cal {L}}_{\rm SG}=\frac{1}{2}(\partial _{\mu }\Phi )^{2}-\frac{m^{2}}{\beta ^{2}}(1-\cos (\beta \Phi )),
\end{equation}
 where \( \Phi (x,t) \) is a scalar field, \( \beta  \) is a real
dimensionless coupling and \( \Phi _{B}(t)=\Phi (x,t)|_{x=0} \). To
preserve the integrability of the bulk theory the boundary potential
is chosen as \[
V_{\rm B}(\Phi _{B})=M_{0}\left( 1-\cos \left( \frac{\beta }{2}(\Phi _{B}-\phi _{0})\right) \right) ,\]
 where \( M_{0} \) and \( \phi _{0} \) are free parameters. As a
result the scalar field satisfies the boundary
condition: 
\begin{equation}
\label{hatfel}
\partial _{x}\Phi |_{x=0}=-M_{0}\frac{\beta }{2}
\sin \left( \frac{\beta }{2}(\Phi _{B}-\phi _{0})\right) .
\end{equation}
Collecting all the possible equivalences between the boundary parameters
their fundamental domain turns out to be \cite{patr} \cite{genpap}:
\[
0\leq M_{0}\leq \infty \quad ;\qquad 0\leq \phi _{0}\leq \frac{\pi
}{\beta }\,.\]

In
the classical theory the two static solutions with lowest energy are given 
 by a static bulk soliton/antisoliton \lq standing
at the right place'  \cite{SSW} \cite{genpap}: 
i.e. by choosing \( \Phi \equiv \Phi _{s}(x,a^+) \)
or \( \Phi \equiv \Phi _{\bar{s}}(x,a^-) \) for \( x\leq 0 \),
where \[
\Phi _{s}(x,a^+)=\frac{4}{\beta }{\textrm{arctg}}(e^{m(x-a^+)}),\qquad \qquad \Phi _{\bar{s}}(x,a^-)=\frac{2\pi }{\beta }-\Phi _{s}(x,a^-),\]
 and \( a^\pm \)  are determined by the boundary condition, eq.(\ref{hatfel}):
 \[
\sinh (ma^\pm)=\frac{\frac{4m}{M_{0}\beta ^{2}}\pm\cos (\frac{\beta }{2}\phi _{0})}{\sin (\frac{\beta }{2}\phi _{0})}.\]
 ($a^+$ and $a^-$ are obtained from each other by \( \phi _{0}\leftrightarrow \frac{2\pi }{\beta }-\phi _{0} \)).
The energies of these two solutions can be written as \begin{eqnarray}
E_{s}(M_{0},\phi _{0}) & \equiv  & E_{\rm{bulk}}+V_{\rm B}=\frac{4m}{\beta ^{2}}+M_{0}-M_{0}R(+),\nonumber \\
E_{\bar{s}}(M_{0},\phi _{0}) & = & \frac{4m}{\beta ^{2}}+M_{0}-M_{0}R(-)=E_{s}(M_{0},\frac{2\pi }{\beta }-\phi _{0}),\label{energiak} 
\end{eqnarray}
 where we introduced 
\[ R(\pm )=\left[ 1\pm 2A\cos (\alpha)+A^2\right] ^{1/2},\qquad A=
\frac{4m}{M_{0}\beta ^{2}},\qquad \alpha =\frac{\beta}{2}\phi_{0} . \]
The difference between these two energies, which is called below
 the \lq classical energy difference', 
\[ \Delta E_{\rm cl}\equiv 
E_{\bar{s}}(M_{0},\phi _{0})-E_{s}(M_{0},\phi _{0})=M_0(R(+)-R(-)),\]
is positive  for \(\alpha\in [0,\frac{\pi}{2})\), $M_0>0$ 
showing that in this range the soliton
generates the ground state and the antisoliton the first excited one.
 From eq.(\ref{energiak})
it follows that for \( \phi _{0}\rightarrow 0^+ \)\footnote{This limit
is not smooth, see our remark later.}
\begin{equation}
\label{mcritcl}
E_{s}=0,\qquad \qquad E_{\bar{s}}=\left\{ \begin{array}{c}
2M_{0}\qquad M_{0}<\frac{4m}{\beta ^{2}}\\
\frac{8m}{\beta ^{2}}\qquad M_{0}>\frac{4m}{\beta ^{2}}
\end{array}\right. \, \, .
\end{equation}

In the process of semi-classical quantisation the oscillators
associated to the linearized
fluctuations around the static solutions \(\Phi
(x,t)=\Phi_{s,\bar{s}}+e^{i\omega t}\xi_{\pm} (x)\) are quantised \cite{Raj}. The
equations of motion of these fluctuations can be written:
\be\label{mozg}
\Bigl[ -\frac{d^2}{dx^2}+m^2-\frac{2m^2}{\cosh^2(m[x-a^\pm])}\Bigr] \xi_{\pm}
(x)=\omega^2 \xi_{\pm} (x);\qquad x<0\ ,
\ee
and \(\xi_{\pm} (x)\) must satisfy also the linearized version of the
boundary condition 
(\ref{hatfel}):
\be\label{lhatfel}
\xi_{\pm}^\prime (x)\vert_{x=0}=-\frac{M_0\beta^2}{4}\frac{1\pm
A\cos\alpha}{R(\pm )}\xi_{\pm} (0).
\ee 
These eigenvalue problems can be solved exactly by mapping eq.(\ref{mozg}) to a
hypergeometric differential equation \cite{LL}. 

\subsection{Discrete spectrum}

In case of the {\sl discrete spectrum} it is convenient to write 
\(\omega^2=m^2(1-\epsilon^2)\). The normalizable 
solutions of eq.(\ref{mozg}) must vanish at \(x\rightarrow -\infty\),
and assuming \(\epsilon \) to be positive, they are given by:
\[
\xi_{\pm}(x)=N e^{m\epsilon (x-a^\pm)}(\epsilon -\tanh
[m(x-a^\pm)]).
\]
The boundary conditions, eq.(\ref{lhatfel}), determine the possible
values of \(\epsilon\) as
\[
\epsilon^2+\epsilon\frac{R(\pm )}{A}\pm\frac{\cos\alpha }{A}=0.
\] 
It is easy to show, that for the solitonic ground state there is no
positive solution of this equation, while for the antisolitonic \lq
exited' state one of the roots, namely 
\be
\epsilon=\frac{R(+)-R(-)}{2A},
\ee
is positive. In fact a simple (numerical) study shows that for all
positive $A$-s and \(\alpha\in [0,\frac{\pi}{2})\) 
\[
0\le \frac{R(+)-R(-)}{2A}\le 1,\quad {\rm and}\quad
\frac{R(+)-R(-)}{2A}=1\quad {\rm iff}\ \alpha=0,\ {\rm and}\ A<1.
\]
In the framework of semi-classical quantisation these findings imply,
that there are no boundary bound states for the ground state,
described by $\Phi_s$, while for the state, described by
$\Phi_{\bar{s}}$, there is such a boundary bound state. The
semi-classical energy of this bound state,
\be\label{omeg}
\omega_0=m\sqrt{1-\left(\frac{R(+)-R(-)}{2A}\right)^2},
\ee
is real, \(\omega_0\ge 0\), and it vanishes only for $\alpha=0$ and
$A<1$. In contrast to the traditional zero modes this vanishing  $\omega_0$ has
nothing to do with $\Phi_{\bar{s}}$ not being invariant under a
continuous symmetry of the Lagrangian, and it indicates some sort of
instability of the state described by $\Phi_{\bar{s}}$. Indeed with
this $\alpha$ and $A$ values (\ref{mcritcl}) gives an energy
difference which 
 is precisely the mass of the bulk soliton, and since
topological charge is not conserved in the boundary theory, the higher energy state can decay
into the lower one by emitting a standing soliton.

At this point it is worth comparing the stability analysis of this 
$\alpha\rightarrow 0$ situation and the one when $\alpha =0$ is set
from the start, to emphasize the non smooth nature of the limit. 
In the latter case the two classical solutions become 
$\Phi_1\equiv\frac{2\pi}{\beta}$ and $\Phi_2\equiv 0$. Repeating the
stability analysis reveals that there are no normalizable bound state
solutions of the fluctuation equations for the ground state,
\(\Phi_2\), while for the \lq excited' state, \(\Phi_1\), there is a
normalizable solution with \(\omega^2=m^2(1-A^{-2})\). When $A>1$ 
this solution signals the existence of a boundary state, while for
$A<1$, when this $\omega^2 $ becomes negative, it indicates the
instability of \(\Phi_1\). The instabilities found both in the
$\alpha\rightarrow 0$ and in the $\alpha\equiv 0$ cases are
consistent with the results of the bootstrap solution \cite{patr} \cite{genpap} showing no
excited boundary states in this range of parameters. 
  
\subsection{Continuous spectrum}

In case of the {\sl continuous spectrum} it is convenient to put
$\omega^2=m^2+q^2$ (with $q\ge 0$). Then the solutions of
eq.(\ref{mozg}), which asymptotically become plane waves, can be
written as
\[
\xi_\pm(x)=\tilde{A}_\pm e^{-iq(x-a^\pm )}\frac{iq+m\tanh
(m[x-a^\pm])}{iq+m}+
 \tilde{B}_\pm e^{iq(x-a^\pm )}\frac{iq-m\tanh
(m[x-a^\pm])}{iq-m}.
\]
The ratio \(\tilde{A}_\pm /\tilde{B}_\pm \) is determined by the
boundary condition eq.(\ref{lhatfel}) at $x=0$. Using this value
the asymptotic ($x\rightarrow -\infty$) form of the fluctuations can
be written as
\[
\xi_\pm (x)\rightarrow C_\pm (e^{ixq}+e^{-ixq}e^{i\delta^\pm (q)}),
\]
where the classical reflection factor is
\be\label{klrefl}
e^{i\delta^\pm (q)}=\frac{m-iq}{m+iq}\,\frac{\pm A^{-1}\cos\alpha
-\frac{q^2}{m^2}+i\frac{q}{m}\frac{R(\pm)}{A}}{\mp A^{-1}\cos\alpha
+\frac{q^2}{m^2}+i\frac{q}{m}\frac{R(\pm)}{A}}\ .
\ee
To handle the infinite volume limit it is convenient to confine the
fluctuations to a box of size $L$, (i.e. to limit $x$ to the section
\((-L,0)\)), and impose Neumann boundary conditions at $x=-L$:
$\xi^\prime (-L)=0$. This condition then determines the possible values of the
momenta:
\be\label{kvkond}
q_n^\pm 2L+\delta^\pm (q_n^\pm )=2n\pi,\qquad n\quad {\rm integer}.
\ee

The semi-classical correction to the classical energy difference,
\(\Delta E_{\rm cl}\), is given by the difference between the sums of
the zero point energies of the fluctuations around \(\Phi_{\bar{s}}\)
and \(\Phi_s\):
\[
\Delta E_{\rm semi}=\Delta E_{\rm cl}+\Delta E_{\rm cor}=\Delta E_{\rm
cl}+\frac{\omega_0}{2}+\frac{1}{2}
\sum\limits_n\left(\sqrt{m^2+(q_n^-)^2}-\sqrt{m^2+(q_n^+)^2}\right)\ .
\]
Replacing -- as usual -- the sum over $n$ by an appropriate integral
in the $L\rightarrow\infty$ limit, exploiting the 
\[
q_n^-=q_n^++\frac{\delta^+-\delta^-}{2L}
\]
consequence of eq.(\ref{kvkond}), and dropping all terms vanishing for
$L\rightarrow\infty$ gives:
\be\label{unren}
\eqalign{\Delta E_{\rm semi}=&\Delta E_{\rm cl}+\frac{\omega_0}{2}
-\frac{M_0\beta^2}{8\pi}\left( R(+)-R(-)\right)+\frac{1}{2\pi}\Bigl[ 
\frac{m}{A}\left( R(-)-R(+)\right)I_1\cr
&-\frac{m\cos\alpha}{A^2}\left( R(+)+R(-)\right)I_2\Bigr]\ , }
\ee 
where
\[
\eqalign{I_1=\int\limits_0^\infty
dy&\frac{y^2\sqrt{1+y^2}}{D},\qquad I_2=\int\limits_0^\infty
dy\frac{\sqrt{1+y^2}}{D},\cr
&D=y^4+(1+A^{-2})y^2+A^{-2}\cos^2\alpha\ .}
\]

\subsection{Renormalization}

The first integral in eq.(\ref{unren}) is logarithmically divergent,
showing the need of regularization and renormalization. This is hardly
surprising since neither the bulk nor the boundary potentials are
normal ordered, and already in the classic paper \cite{dhn} it is
shown on the example of the bulk soliton's mass correction, that this
naive procedure leads to logarithmic divergences even in mass differences. The proper way to
deal with these infinities \cite{dhn} \cite{Raj} is to use the
counterterms, that account for the difference between the normal ordered
and non ordered potentials.

In the boundary sine-Gordon model we use the same counterterm for the
bulk potential as in the bulk theory:
\[
V_{\rm count}[\Phi ]=-\frac{\delta
m^2}{\beta^2}\int\limits_{-\infty}^0dx\left(1-\cos(\beta\Phi
)\right);\qquad \delta
m^2=-\frac{m^2\beta^2}{4\pi}\int\limits_0^\Lambda
\frac{dk}{\sqrt{k^2+m^2}};
\]
but the integral is over the \(x\le 0\) half space only. The argument
for this choice is based on its local nature: as such it should be
independent of the presence of the boundary. For the boundary
potential we {\sl assume} that its counterterm has an analogous form
\[
V_{\rm B\ count}[\Phi ]=-\delta
M_0\left(1-\cos(\frac{\beta}{2}(\Phi_{\rm B}-\phi_0))\right),
\]
with $\delta M_0$ being some parameter. The total contribution of
counterterms to the energy difference 
\[
CT=V_{\rm count}[\Phi_{\bar{s}} ]+V_{\rm B\  count}[\Phi_{\bar{s}} ]-
V_{\rm count}[\Phi_s ]- V_{\rm B\ count}[\Phi_s ]
\]
may remove the logarithmic divergence in eq.(\ref{unren}), if it is proportional to
$R(+)-R(-)$. This condition determines \(\delta M_0\):
\[
\delta M_0=-\frac{M_0\beta^2}{4\cdot2\pi}\int\limits_0^\Lambda
\frac{dk}{\sqrt{k^2+m^2}},
\]
and with this choice $CT$ becomes
\[
CT=\frac{m}{2\pi A}\left( R(+)-R(-)\right)\int\limits_0^{\Lambda /m}
\frac{dy}{\sqrt{y^2+1}}.
\]
Since the overall magnitude of $CT$ is fixed by $\delta m^2$ there are
no more free parameters. Thus the fact that adding $CT$ to $\Delta E_{\rm
semi}$ {\sl does} remove the divergence gives a partial justification of the
renormalization procedure used.\footnote{By setting up a systematic
perturbation theory in boundary sine-Gordon model treating
simultaneously both the bulk and the boundary interactions one can
confirm the correctness of both $\delta m^2$ and $\delta M_0$
\cite{Zoli}.} In the renormalized energy difference
\[
\Delta E^{\rm ren}_{\rm semi}=\Delta E_{\rm semi}+CT
\]
only the term containing $I_1$ gets modified and is replaced by
\[
\frac{m}{2A\pi}\left( R(-)-R(+)\right)I_1\quad\rightarrow\quad 
 \frac{m}{2A^3\pi}\left( R(+)-R(-)\right)\tilde{I}_1,
\]
with
\[
\tilde{I}_1=\int\limits_0^\infty\frac{dy}{\sqrt{1+y^2}}\,\frac{y^2+\cos^2\alpha}{D}.
\]
The convergent integrals $\tilde{I}_1$, and $I_2$ can be computed
symbolically with the aid of Maple. For this it is helpful to write
$D=(y^2+a)(y^2+b)$ with
\[
a=\left(\frac{R(+)+R(-)}{2A}\right)^2\ge 1,\qquad 
b=\left(\frac{R(+)-R(-)}{2A}\right)^2,\quad 0\le b\le 1,
\]
and tell Maple the range of these parameters. Using the explicit form
of these integrals, after some algebra, the renormalized energy
difference is obtained as
\be\label{ren}
\eqalign{\Delta E^{\rm ren}_{\rm semi}=&M_0(R(+)-R(-))+
\frac{m}{2}\sqrt{1-\left(\frac{R(+)-R(-)}{2A}\right)^2}
-\frac{M_0\beta^2}{8\pi}\left( R(+)-R(-)\right) \cr
&-\frac{m}{\pi} \sqrt{1-\left(\frac{R(+)-R(-)}{2A}\right)^2}
{\rm{arccos}}\left(\frac{R(+)-R(-)}{2A}\right) \ .} 
\ee
It is a remarkable feature of this expression, that it depends only on
the difference \newline $\left( R(+)-R(-)\right)/(2A)$.

\section{Comparison to the exact results}

In this section the main results of the previous 
semi-classical quantisation,
namely the (non) existence of semi-classical bound states, the classical 
reflection factors and the semi-classically corrected energy difference
are compared to the results obtained from the exact (bootstrap)
solution. 

In this process the sine-Gordon field is assumed to correspond to  
the semi-classical
limit of the first breather, while the exact ground state \(\vert
\rangle\) and the first excited boundary state \(\vert 0\rangle\) are
identified as the quantum analogues of the classical states (solutions) 
\(\Phi_s\), \(\Phi_{\bar{s}}\).  This latter
identification was suggested in \cite{genpap}
on the basis of the existence of a  ($Z_2$ 
reflection type) transformation that changes the roles of these two
states in the same way as the classical 
\(\Phi\leftrightarrow\frac{2\pi}{\beta}-\Phi\), 
\(\phi_0\leftrightarrow\frac{2\pi}{\beta}-\phi_0\) changes \(\Phi_s\)
and \(\Phi_{\bar{s}}\) into each other.

In the exact solution of the boundary sine-Gordon model \cite{GZ},
\cite{patr}, 
\cite{genpap}, \cite{gosh} the coupling constant $\beta $ appears
through 
\[
 \lambda =\frac{8\pi}{\beta^2}-1,
\]
while the dependence on the boundary condition appears in the form of
two real parameters, \(\eta\) and \(\vartheta\), the fundamental ranges
of which are \cite{genpap}
\[
 0\le\eta\le\frac{\pi}{2}(\lambda +1),\qquad 0\le\vartheta\le\infty\ .
\]
Boundary bound states appear in the exact solution as poles in the
various reflection amplitudes at purely imaginary rapidity
\(u=-i\theta\). The location of these 
poles depends on the $\eta$ parameter only and 
is given by    appropriate combinations of
\[
 \nu_n=\frac{\eta}{\lambda}-(2n+1)\frac{\pi}{2\lambda},\qquad 
  w_k=\frac{\bar{\eta}}{  \lambda}-(2k+1)\frac{\pi}{2\lambda},\qquad
\bar{\eta}=\pi(\lambda +1)-\eta \ .
\]
Though the semi-classical quantisation is non perturbative, its validity
is restricted to weak coupling \cite{Raj}, which in our case means to 
$\beta\rightarrow 0$. Therefore it is the $\lambda\rightarrow\infty$
limit of the exact solution that should be compared to the semi-classical 
results. The $\eta$ parameter should be scaled to obtain a non trivial 
spectrum in this limit, and we propose to write
\[
 \eta=c\frac{\pi}{2}(\lambda +1), \qquad 0\le c\le 1,
\]
and keep $c$ fixed.

\subsection{Boundary states}

The reflection factor of the first breather, \(B^1\), on the ground
state boundary is given by \cite{gosh}
\begin{equation}
\label{b1_refl}
R^{(1)}(\theta )=\frac{\left( \frac{1}{2}\right) \left(
\frac{1}{2\lambda }+1\right) }{\left( \frac{1}{2\lambda
}+\frac{3}{2}\right) }\frac{\left( \frac{\eta }{\pi \lambda
}-\frac{1}{2}\right) \left( \frac{i\vartheta }{\pi \lambda
}-\frac{1}{2}\right) }{\left( \frac{\eta }{\pi \lambda
}+\frac{1}{2}\right) \left( \frac{i\vartheta }{\pi \lambda
}+\frac{1}{2}\right) }\quad ,\quad 
 (x)=\frac{\sinh \left( \frac{\theta
}{2}+i\frac{\pi x}{2}\right) }{\sinh \left( \frac{\theta
}{2}-i\frac{\pi x}{2}\right) }\,.
\end{equation}
($\theta $ is the rapidity of $B^1$). 
$B^1$'s reflection factor on \(\vert 0\rangle \), 
\(R^{(1)}_{\vert 0\rangle}(\theta )\),  
is obtained from this 
expression by the substitution 
\(\eta\rightarrow\bar{\eta}=\pi (\lambda +1)-\eta\) \cite{genpap} (see
also \cite{patr}).
The only pole of \(R^{(1)}(\theta )\) which may describe a boundary
state is at
\[
 \frac{\eta}{\lambda}-\frac{\pi}{2}=\frac{1}{2}(\nu_0-w_1)\ .
\]
This corresponds to a bound state if it is in the physical strip, i.e. if 
 \(0\le\frac{1}{2}(\nu_0-w_1)\le\frac{\pi}{2} \). In the semi-classical 
($\lambda\rightarrow\infty$) limit, keeping $c$ fixed,
\[
 \frac{1}{2}(\nu_0-w_1)=(c-1)\frac{\pi}{2}+\frac{c\pi}{2\lambda}\sim 
(c-1)\frac{\pi}{2}\ ,
\] 
and since this is negative we conclude that \(B^1\) can not create a
bound state on \(\vert\rangle\). On the other hand, 
 \(R^{(1)}_{\vert 0\rangle}(\theta )\) has a pole at
\[
 \frac{\pi}{\lambda}-\frac{\eta}{\lambda}+\frac{\pi}{2}=
\frac{1}{2}(w_0-\nu_1)\ ,
\]  
which may describe a bound state 
if it is in the physical strip. Since in the semi-classical limit 
\[
 \frac{1}{2}(w_0-\nu_1)=(1-c)\frac{\pi}{2}+\frac{(2-c)\pi}{2\lambda}\sim 
(1-c)\frac{\pi}{2}\ 
\] 
is in the physical strip we conclude that \(B^1\) can create a bound
state (in fact it is the state $\vert 1\rangle$) when reflecting 
on \(\vert 0\rangle\). Recalling, that semi-classically $B^1$ should
correspond to the sine-Gordon field, we see that
these findings fit nicely with the semi-classical
results and strengthen the association 
\((\Phi_s\,,\Phi_{\bar{s}})\leftrightarrow (\vert\rangle\,,\vert
0\rangle )\). 

The energy of this bound state above \(E_{\vert 0\rangle}\) is
determined by the location of the pole
\be\label{b1dif}
E-E_{\vert 0\rangle}=m_1\cos\left((1-c)\frac{\pi}{2}
+\frac{(2-c)\pi}{2\lambda} \right)\ ,
\ee
where $m_1=2M\sin\left(\frac{\pi}{2\lambda}\right)$ is the mass of the 
$B^1$ and $M$ is the soliton mass. Using the semi-classical expression 
\(M=\frac{8m}{\beta^2}\left(1-\frac{\beta^2}{8\pi}\right)\) one finds
from (\ref{b1dif}) for \(\lambda\rightarrow\infty\) (\(\beta\rightarrow
0\))
\[
 E-E_{\vert 0\rangle}\sim m\sin\left(\frac{c\pi}{2}\right)\ .
\]
Identifying this limiting energy difference with the energy of the
semi-classical bound state $\omega_0$, eq.(\ref{omeg}), determines the
(limiting value of the) \lq infrared' (bootstrap) 
parameter $\eta$ in terms of the 
\lq ultraviolet' (Lagrangian) $M_0$ and $\phi_0$:
\be\label{id1}
\sin\left(\frac{c\pi}{2}\right)=
\sqrt{1-\left(\frac{R(+)-R(-)}{2A}\right)^2}\,.
\ee

\subsection{The limit of the reflection factors}

The next step is to establish a relation between the (semi)classical limits of $R^{(1)}(\theta
)$ and \(R^{(1)}_{\vert 0\rangle}(\theta )\), and the classical
reflection factors \(e^{i\delta^\pm (q)}\). Since the exact quantum
reflection factors  eq.(\ref{b1_refl})  depend also on the $\vartheta$
 parameter, for a non trivial limit we have to scale also this parameter. In
analogy with the $\eta$ parameter we propose to write
\[
\vartheta=\vartheta_{\rm cl}(\lambda +1),\qquad 0\le\vartheta_{\rm
cl}\le\infty .
\]
This way, keeping only the leading constant terms in the
$\lambda\rightarrow\infty$ limit, one obtains:
\be\label{reflim}
R^{(1)}(\theta )\rightarrow\frac{i\sinh\theta -1}{i\sinh\theta +1}\,
\frac{\cos\left(\frac{c\pi}{2}\right)\cosh\vartheta_{\rm
cl}-\sinh^2\theta +i\sinh\theta\left(\cos\left(\frac{c\pi}{2}\right)+ 
\cosh\vartheta_{\rm cl}\right)}    
 {\cos\left(\frac{c\pi}{2}\right)\cosh\vartheta_{\rm
cl}-\sinh^2\theta -i\sinh\theta\left(\cos\left(\frac{c\pi}{2}\right)+ 
\cosh\vartheta_{\rm cl}\right)}\,.
\ee 
The expression for the limiting value of \(R^{(1)}_{\vert
0\rangle}(\theta )\) is obtained by making the substitution
\(c\rightarrow\bar{c}=2-c\), (which amounts to changing the sign of 
\(\cos\left(\frac{c\pi}{2}\right)\)) in eq.(\ref{reflim}). 
Identifying these
limiting $R^{(1)}(\theta
)$ and \(R^{(1)}_{\vert 0\rangle}(\theta )\) with
\(e^{i\delta^\pm (q)}\), eq.(\ref{klrefl}), using \(\sinh\theta=\frac{q}{m}\), determines
the bootstrap parameters \(\frac{c\pi}{2}\) and \(\vartheta_{\rm cl}\)
as
\be\label{egye}
\eqalign{\cos\left(\frac{c\pi}{2}\right)+ 
\cosh\vartheta_{\rm cl}&=\frac{R(+)}{A},\cr
\cosh\vartheta_{\rm cl}-\cos\left(\frac{c\pi}{2}\right) &=\frac{R(-)}{A},}
\ee
together with
\be\label{egyk}
\cos\left(\frac{c\pi}{2}\right)\cosh\vartheta_{\rm cl}=\frac{\cos\alpha}{A}.
\ee
The algebraic solution of eq.(\ref{egye}) 
\be\label{ameg}
\cos\left(\frac{c\pi}{2}\right)=\frac{R(+)-R(-)}{2A},\qquad 
\cosh\vartheta_{\rm cl}=\frac{R(+)+R(-)}{2A},
\ee
satisfies eq.(\ref{egyk}) and is also consistent with eq.(\ref{id1}). 

\subsection{The limit of $E_{\vert 0\rangle}-E_{\vert\rangle}$ and the
UV-IR relation}

According to the bootstrap solution \cite{patr} \cite{genpap} the energy difference
between the lowest excited boundary state and the ground state is
given by
\[
\Delta E_{\rm bst}\equiv E_{\vert
0\rangle}-E_{\vert\rangle}=M\cos\nu_0=M\cos\left(\frac{\eta}{\lambda}-
\frac{\pi}{2\lambda}\right),
\]
where $M$ is the soliton mass. In the semi-classical limit, using the
appropriately scaled $\eta$ parameter, this can be written as
\be\label{egveg}
\Delta E_{\rm bst}=M\cos\left(\frac{c\pi}{2}\right)-
M\sin\left(\frac{c\pi}{2}\right)\frac{\beta^2}{8\pi}\left(\frac{c\pi}{2}-\frac{\pi}{2}\right)
+M{\cal O}(\beta^4)\,.
\ee
Now it is easy to show,
using the complete semi-classical expression, 
\(M=\frac{8m}{\beta^2}\left(1-\frac{\beta^2}{8\pi}\right)\),
in the first term, the leading \(M=\frac{8m}{\beta^2}\) in the (higher
order) second one, together with the actual value of
\(\cos\left(\frac{c\pi}{2}\right)\) in (\ref{ameg}), that the first
four terms of $\Delta E_{\rm bst}$  coincide term by term  with the expression of
$\Delta E^{\rm ren}_{\rm semi}$ eq.(\ref{ren}). 

Now we can understand the importance of the fact 
 that in spite of the
intermediate stages the dependency on \((R(+)+R(-))/(2A)\) cancels
in the final form of 
the semi-classical
$\Delta E^{\rm
ren}_{\rm semi}$. This should happen since \(\Delta E_{\rm bst}\),     
just as the whole spectrum of boundary
states predicted by the bootstrap solution,  
is also independent of $\vartheta$
 thus  
in the semi-classical limit it should depend only on $\frac{c\pi}{2}$
but should be independent of $\vartheta_{\rm cl}$. 

The nice matching between  $\Delta E^{\rm
ren}_{\rm semi}$ and \(\Delta E_{\rm bst}\) confirms the relation
between the bootstrap and Lagrangian parameters eq.(\ref{ameg}). This
relation makes it possible to determine the (semi-classical limit of
the) only free parameter in the
so called UV-IR relation.

On general grounds the generic form of the relation between the
bootstrap and Lagrangian parameters of boundary sine-Gordon model (i.e
of the UV-IR relation) is
\be\label{uvirr}
\eqalign{\cos\left(\frac{\eta}{\lambda +1}\right) 
\cosh\left(\frac{\vartheta}{\lambda +1}\right)&=\frac{M_0}{M_{\rm crit}}\cos\alpha\,,\cr
\sin\left(\frac{\eta}{\lambda +1}\right) 
\sinh\left(\frac{\vartheta}{\lambda +1}\right)&=\frac{M_0}{M_{\rm
crit}}\sin\alpha\,,}
\ee
where the parameter $M_{\rm crit}$ ($M_0/M_{\rm crit}$) 
may depend on \(\beta\). Our
aim is to say something on this parameter and on this 
dependence. First of all, 
\(\frac{\eta}{\lambda +1}\) and \( \frac{\vartheta}{\lambda +1}\) are
nothing but $c\pi /2$ and $\vartheta_{\rm cl}$ in the way they were
introduced, thus eq.(\ref{uvirr}) determines in fact 
these parameters {\sl for all values
of} $\lambda $. Making this identification explicit 
in eq.(\ref{uvirr}) and comparing
to eq.(\ref{egyk}) gives, that in the semi-classical limit
\be\label{kvmcr}
\frac{M_0}{M_{\rm crit}}=\frac{1}{A},\qquad\quad {\rm i.e.}\qquad\quad
M_{\rm crit}=\frac{4m}{\beta^2}\ .
\ee
Note that this is the same value as the classical one appearing in
eq.(\ref{mcritcl}). 

There are several points that should be stressed about \(M_{\rm crit}\) 
in general and its actual value in particular. The first point to
mention is that $M_0/M_{\rm crit}$ appearing in eq.(\ref{uvirr}) 
may depend on the regularization scheme used to define the quantum
theory and the value in (\ref{kvmcr}) is in the \lq semi-classical
scheme'. In a recent paper Corrigan and Taormina obtained the UV-IR
relation in sinh-Gordon model by semi-classically quantising the
(periodic) boundary breathers \cite{CT}. Analytically continuing their
results in $\beta$ (and accounting for the differences between the
parameters) one can show, that their \(M_{\rm crit}\) is identical to 
eq.(\ref{kvmcr}). In this respect it is worth emphasizing that the
analogues of the static solutions $\Phi_s$ and $\Phi_{\bar{s}}$ just
like the states $\vert\rangle$ and $\vert 0\rangle$, 
upon which our investigation is based, are {\sl absent} in
the sinh-Gordon theory, thus the results of this paper give an
independent confirmation of the  \(M_{\rm crit}\) obtained in
\cite{CT}. 

In \cite{CT} it is conjectured that this result for 
\(M_{\rm crit}\) may be exact. To support this conjecture we note that
our results make it possible to check that \(M_{\rm crit}\) receives no 
${\cal O}(\beta^0)$ correction:
\[
M_{\rm crit}=\frac{4m}{\beta^2}\left(1+{\cal O}(\beta^4)\right)\,.
\]  
To show this denote the ($\beta$ dependent) 
$M_0/M_{\rm crit}$ as $H$  and determine  
$\cos\left(\frac{c\pi}{2}\right)$ from eq.(\ref{uvirr})
\[
 \cos\left(\frac{c\pi}{2}\right)=\frac{H}{2}
\left(\sqrt{1+H^{-2}+2H^{-1}\cos\alpha}-\sqrt{1+H^{-2}-2H^{-1}\cos\alpha} \right)\ ,
\]  
and finally write $H=\frac{1}{A}\left(1+\delta
H\frac{\beta^2}{8\pi}\right)$. Now plugging this expression for 
$\cos\left(\frac{c\pi}{2}\right)$ (and the equivalent one for
$c\pi/2$) into (\ref{egveg}) reveals
that the only choice that  guarantees the agreement 
between eq.(\ref{egveg}) and eq.(\ref{ren}) is 
$\delta H=0$.\footnote{Since the 
$M{\cal O}(\beta^4)$ terms are not calculated we cannot say anything about
the higher order corrections.}  

Perturbed conformal field theory is another useful scheme to describe
the boundary sine-Gordon model. In this description BSG is viewed as a
$c=1$ boundary CFT perturbed by the (relevant) vertex operators
constituting the bulk and boundary potentials \cite{neupap}:
\[
S=S_{c=1}+\frac{\mu}{2}\int\limits_{-\infty}^\infty
dt\int\limits_{-\infty}^0dx(V_\beta [\Phi ]+V_{-\beta}[\Phi ])+
\frac{\tilde{\mu}}{2}\int\limits_{-\infty}^\infty
dt(\Psi_{\beta /2} [\Phi ]{\rm e}^{-i\alpha}+
\Psi_{-\beta /2} [\Phi ]{\rm e}^{i\alpha})\,,
\]
where
\[
V_\beta [\Phi ]=n(z,\bar{z}):{\rm e}^{i\beta\Phi (x,t)}:\,,\qquad
\Psi_{\beta /2}[\Phi ]=:{\rm e}^{i\frac{\beta}{2}\Phi (0,t)}:\,,
\]
and $n(z,\bar{z})$ denotes the appropriate normal ordering
function. The $\mu $ and $\tilde{\mu}$ parameters play the role of $m$
and $M_0$ respectively and have non trivial dimensions:
\[
[\mu ]={\rm mass\ }^{2-\frac{\beta^2}{4\pi}},\qquad 
[\tilde{\mu} ]={\rm mass\ }^{1-\frac{\beta^2}{8\pi}}.
\]
The relation between $\mu$ and the soliton mass $M$ is known from a
TBA study of the bulk sine-Gordon model \cite{Zamm}
\be\label{muM}
\mu =\kappa (\beta)M^{2-2\Delta},\qquad 
\kappa (\beta)=\frac{2\Gamma(\Delta)}{\pi\Gamma(1-\Delta)}
\left(\frac{\sqrt\pi\Gamma\left(\frac1{2-2\Delta}\right)}
{2\Gamma\left(\frac{\Delta}{2-2\Delta}\right)}\right)
^{2-2\Delta},\quad \Delta=\frac{\beta^2}{8\pi}.
\ee
In this scheme the UV-IR relation takes the form of eq.(\ref{uvirr})
with the replacement
\be\label{zrel}
\frac{M_0}{M_{\rm crit}}\rightarrow\frac{\tilde{\mu}}{\mu_{\rm crit}},
\qquad \mu_{\rm crit}=\sqrt{\frac{2\mu}{\sin\frac{\beta^2}{8}}}.
\ee
This relation was obtained by Al.B. Zamolodchikov \cite{Zupb} and has recently
been verified by a TCSA study of the spectrum of boundary sine-Gordon
model \cite{uvir}.

Thus the $\beta$ dependence of the constant on the right hand side of
eq.(\ref{uvirr}) is different in the semi-classical and in the perturbed
CFT schemes. Nevertheless in the semi-classical limit the two results
coincide. In the perturbed CFT scheme the limiting values of $c\pi /2$
and $\vartheta_{\rm cl}$ should be obtained from eq.(\ref{uvirr}) with
$\frac{\tilde{\mu}}{\mu}=H$. Furthermore, for the comparison, the $\mu$, $\tilde{\mu}$ and
the $m$, $M_0$ parameters of the two schemes should be related to each
other. Using the semi-classical expression for $M$ in the
$\beta\rightarrow 0$ limit of eq.(\ref{muM}) gives 
$\mu\rightarrow\frac{m^2}{\beta^2}$  and matching the
leading (classical) term of eq.(\ref{egveg}) to the scheme independent
$\Delta E_{\rm cl}$ 
fixes $\tilde{\mu}\rightarrow M_0$; thus 
$\mu_{\rm crit}\rightarrow\frac{4m}{\beta^2}=M_{\rm crit}$ indeed. 

\section{Semi-Classical soliton reflections}

In this section the semi-classical limits of soliton/antisoliton
reflection amplitudes on the boundary at $x=0$ are studied. The
relevant classical solutions are time dependent - as opposed to the
static ones considered in section 2 - but just like the static ones
are specific to sine-Gordon and have no analogues in sinh-Gordon
theory. A long time ago a completely general 
expression for the semi-classical phase shift was given in
terms of the classical time delay and of the number of semi-classical
bound states by Jackiw and Woo \cite{JW}. The idea to compare in
boundary sine-Gordon model this
expression and the semi-classical limit of the exact reflection
amplitudes (obtained from the bootstrap) as a consistency check and to gain information on the
relation between the Lagrangian and the bootstrap parameters 
was put forward by Saleur, Skorik and Warner (SSW) in \cite{SSW}. SSW
determined the classical time delay in case of soliton/antisoliton reflections
on ground state boundary for the general boundary conditions, but only
for Dirichlet boundary conditions 
made the comparison with the exact results. In this section the
comparison is made in case of ground state boundaries with general
boundary conditions and also for the lowest excited boundary in case of Neumann
boundary condition.  
  
\subsection{Neumann boundary condition}

The expression given in \cite{JW} for the semi-classical phase shift
${\rm e}^{i\delta (E)}$ is
\be\label{JaW}
\delta (E)=n_B\pi +\int\limits_{E_{\rm th}}^EdE^\prime\Delta
t(E^\prime),
\ee
where $n_B$ is the number of the (semi-classical) bound states and
$\Delta t(E^\prime)$ is the classical time delay. As an illustration
consider the (anti)solitons reflecting on a ground state Neumann
boundary, i.e. when $\partial_x\Phi\vert_{x=0}=0$ (corresponding to
$M_0=0$)\footnote{Since the vanishing $M_0$ makes $\alpha$ a redundant
parameter, and the bootstrap parameters take fixed values ($\eta$ 
becomes the maximally allowed $\frac{\pi}{2}(\lambda
+1)$ and $\vartheta $ vanishes) this illustration may serve only as a
consistency check.}.  
Then there are
classical solutions only for solitons reflecting as antisolitons (and
vice versa) but not for solitons reflecting as solitons. Furthermore,
the classical solution describing an asymptotic soliton with velocity
$v$ heading to and reflecting from the boundary at $x=0$ can be
obtained by restricting to the $x\le 0$ half line a special solution
of the bulk theory, that describes a soliton with velocity $v$
scattering on an antisoliton with velocity $-v$ \cite{SSW}, \cite{neupap}.   
Therefore the classical time delay of the soliton reflecting on the
Neumann boundary is identical to the time delay in the soliton
antisoliton scattering in the bulk theory:
\[
\Delta t=\frac{2\ln v}{m\gamma
v}\,,\qquad\gamma=\frac{1}{\sqrt{1-v^2}}\,.
\]
The number of bound states, i.e. the number of boundary breathers with
Neumann b.c. were obtained in \cite{neupap} by semi-classically
quantising the classical boundary breathers with the result that 
$n_B=\left[\frac{\lambda}{2}\right]$. In the semi-classical limit
$\lambda\rightarrow\infty$ thus
$n_B\sim\frac{\lambda}{2}=\frac{4\pi}{\beta^2}$. Since the energy of
the reflecting soliton is 
$E=\frac{M}{\sqrt{1-v^2}}=M\cosh(\theta
)=\frac{8m}{\beta^2\sqrt{1-v^2}}$, eq.(\ref{JaW}) yields in this case
\[
\delta
(E)=\frac{4\pi^2}{\beta^2}+\frac{16}{\beta^2}\int\limits_0^{\tanh\theta}dv^\prime
\frac{\ln v^\prime}{1-v^{\prime 2}}\,.
\]
In the exact solution of BSG with Neumann b.c. there are two
amplitudes that describe the reflections of solitons and antisolitons
on the ground state boundary: $P(\theta )$ describes the \lq diagonal' scattering,
i.e. when solitons reflect as solitons and antisolitons as
antisolitons, while $Q(\theta )$ describes the \lq non - diagonal'
scattering, when solitons reflect as antisolitons (and vice versa). In
\cite{neupap} simple integral representations were given for them:
\[
P(\theta )=
\frac{\sin
(\frac{\lambda\pi}{2})}{\sin\left(\frac{\lambda\pi}{2}+i\lambda\theta\right)}{\rm
e}^{-iI(\lambda ,\theta )},
\qquad
Q(\theta )=-i
\frac{\sinh
(\lambda\theta)}{\sin\left(\frac{\lambda\pi}{2}+i\lambda\theta\right)}{\rm
e}^{-iI(\lambda ,\theta )}\,,
\]
\[
I(\lambda ,\theta)=\int_0^{\infty}\frac{dt}{t}t\sin\left(\frac{2\theta
t}{\pi}\right)
\left[\frac{2\sinh\left(\frac{3t}2\right)\sinh\left(\frac{\lambda-1}{2\lambda}t\right)}
{\sinh\left(\frac{t}{2\lambda}\right)\sinh(2t)}+\frac{\sinh(t/\lambda
)-\sinh(t)}{\cosh(t)\sinh(t/\lambda )}\right].
\]
In the semi-classical limit $P(\theta )\sim{\rm e}^{-\lambda\theta}{\rm
e}^{-iI(\lambda ,\theta )}\rightarrow 0$, which is consistent with the
absence of diagonal classical reflection. On the other hand 
\be\label{qlim}
Q\rightarrow{\rm e}^{i\frac{\lambda\pi}2}{\rm
e}^{-iI_1(\lambda ,\theta )},\quad I_1(\lambda ,\theta
)=\lim_{\lambda\to\infty}I(\lambda
,\theta)=\lambda\int\limits_0^\infty
\frac{dt}{t^2}\sin\left(\frac{2\theta
t}{\pi}\right)\tanh\left(\frac{t}2\right)+{\cal O}(\lambda^0)\,,
\ee
where we neglected all ${\cal O}(\lambda^0)$ terms in the exponents. 
The integral $\partial_\theta I_1$ can be found in Gradstein Ryzhikh,
\cite{GR}, thus
\[
I_1=-\frac{2\lambda}{\pi}\int_0^\theta dv\ln\tanh v=
-\frac{2\lambda}{\pi}\int_0^{\tanh\theta} dv^\prime\frac{\ln
v^\prime}{1-v^{\prime 2}}\,.
\]
Using finally the semi-classical relation
$\lambda\sim\frac{8\pi}{\beta^2}$ in eq.(\ref{qlim}) reproduces the
semi-classical phase shift indeed.

\subsubsection{Excited Neumann boundary}

The exact soliton/antisoliton reflection amplitudes are known also
when the Neumann boundary is in its excited states $\vert n\rangle $
$n=1,\dots ,\left[\frac{\lambda}{2}\right]$\footnote{For Neumann
boundary condition the pole described by $\nu_0$ is at $\theta
=i\frac{\pi}{2}$, and it corresponds to the emission of a
soliton/antisoliton by the boundary \cite{GZ} rather than to a bound
state. 
Alternatively one can say that 
 $\vert 0\rangle$ becomes
identical to the ground state $\vert\rangle$, as not only their
energies but also the $P(\theta )$ and $Q(\theta )$ reflection factors
on them become identical \cite{neupap}.}. The $P$, $Q$ reflection
factors on the lowest excited state $\vert 1\rangle $ change as
\cite{neupap}
\[
P\rightarrow\tilde{P}=P(\theta )B(\lambda ,\theta ),\qquad  
Q\rightarrow\tilde{Q}=Q(\theta )B(\lambda ,\theta ),
\]
\[
B(\lambda
,\theta)=\tan\left[\frac{u}2+\frac{\pi}2\left(\frac{1}\lambda +\frac12\right)\right]
\tan\left[\frac{u}2-\frac{\pi}2\left(\frac{1}{\lambda}-\frac12\right)\right]
\tan^2\left(\frac{u}2+\frac{\pi}4\right),\quad u=-i\theta\,.
\]
In the semi-classical limit 
\[
\lim_{\lambda\to\infty}B(\lambda ,\theta )
=\frac{1-i\sinh\theta}{1+i\sinh\theta}
\tan^2\left(\frac{-i\theta}2+\frac{\pi}4\right),
\]
which gives only an ${\cal O}(\lambda^0)$ correction in the exponent
of $\tilde{Q}$. Thus the leading term in the exponent, i.e. the
semi-classical phase shift, is identical to what was found
for the ground state boundary.
    
With Neumann b.c. the state $\vert 1\rangle$ may be thought of
classically as a (classical) breather bound to the boundary at $x=0$
\cite{neupap}. Thus the classical reflection process may be described
as a soliton antisoliton pair reflecting on the breather at $x=0$, and
the classical time delay should be obtained from this picture. The
relevant classical solution is constructed by the $\tau$ function
method \cite{SSW} \cite{Abl} in two steps. First a 4 soliton solution
describing two pairs of solitons and antisolitons is determined and
the relevant time delays are obtained. Then we continue the parameters
of one of the pairs to purely imaginary values to describe the
breather and make the necessary changes in the expression of the time
delay.

In the $\tau$ function method each soliton and antisoliton is
characterized by its velocity, by its \lq rapidity type' parameter and
by its \lq position type' parameter. In the solution below the
following parameters are used: the soliton of the first (second) pair
moves with velocity $u$ ($v$), its rapidity type parameter is denoted
by $k$ ($p$) and its position type parameter by $a_1$ ($b_1$); for the
antisoliton of the first (second) pair the corresponding quantities
are $-u$ ($-v$), $1/k$ ($1/p$), and $a_2$ ($b_2$) respectively. (These
quantities give a redundant characterization as $u$ and $k$
-alternatively $v$ and $p$ - can be expressed in terms of the
$\theta_1$ and $\theta_2$ rapidities of the first and second solitons:
$u=\tanh\theta_1$, $k={\rm e}^{\theta_1}$;
  $v=\tanh\theta_2$, $p={\rm e}^{\theta_2}$). Then,
using also the
\[
 \gamma=\frac{1}{\sqrt{1-u^2}}\,,\qquad\qquad\tilde{\gamma}=\frac{1}{\sqrt{1-v^2}}
\]
quantities, in the centre of mass system the $\tau$ function of the 
solution may be written
as
$
%\begin{multline}
\tau =1 + e^{-2\gamma x}e^{-a_1-a_2}u^2 - e^{-2\tilde{\gamma} x}e^{-b_1-b_2}u^2 \\
-e^{-\gamma(x+ut)}e^{-\tilde{\gamma}(x+vt)}e^{-a_1-b_1}\left(\frac{k-p}{k+p}\right)^2
+e^{-\gamma(x+ut)}e^{-\tilde{\gamma}(x-vt)}e^{-a_1-b_2}\left(\frac{k-\frac{1}{p}}{k+\frac{1}{p}}\right)
^2\\
+e^{-\gamma(x-ut)}e^{-\tilde{\gamma}(x+vt)}e^{-a_2-b_1}\left(\frac{\frac{1}{k}-p}{\frac{1}{k}+p}\right)
^2-
e^{-\gamma(x-ut)}e^{-\tilde{\gamma}(x-vt)}e^{-a_2-b_2}\left(\frac{\frac{1}{k}-\frac{1}{p}}{\frac{1}{k}+
\frac{1}{p}}\right)^2\\
+e^{-2\gamma x}e^{-2\tilde{\gamma}
x}e^{-a_1-a_2-b_1-b_2}u^2v^2\left(\frac{k-p}{k+p}\right)^2\left(\frac{k-\frac{1}{p}}{k+\frac{1}{p}}\right)^2\left(\frac{\frac{1}{k}-p}{\frac{1}{k}+p}\right)^2\left(\frac{\frac{1}{k}-\frac{1}{p}}{\frac{1}{k}+\frac{1}{p}}\right)^2\\
{+i\bigl[ e^{-\gamma(x+ut)}e^{-a_1}
-e^{-\gamma(x-ut)}e^{-a_2} +e^{-\tilde{\gamma}(x+vt)}e^{-b_1}-
e^{-\tilde{\gamma}(x-vt)}e^{-b_2}}\\ 
{+e^{-2\gamma
x}e^{-\tilde{\gamma}(x+vt)}e^{-a_1-a_2-b_1}u^2\left(\frac{k-p}{k+p}\right)^2\left(\frac{\frac{1}{k}-p}{\frac{1}{k}+p}\right)^2}
{-e^{-2\gamma
x}e^{-\tilde{\gamma}(x-vt)}e^{-a_1-a_2-b_2}u^2\left(\frac{k-\frac{1}{p}}{k+\frac{1}{p}}\right)^2\left(\frac{\frac{1}{k}-\frac{1}{p}}{\frac{1}{k}+\frac{1}{p}}\right)^2}\\
{+e^{-2\tilde{\gamma}
x}e^{-\gamma(x+ut)}e^{-a_1-b_1-b_2}v^2\left(\frac{k-p}{k+p}\right)^2\left(\frac{k-\frac{1}{p}}{k+\frac{1}{p}}\right)^2}
-e^{-2\tilde{\gamma}
x}e^{-\gamma(x-ut)}e^{-a_2-b_1-b_2}v^2\left(\frac{\frac{1}{k}-p}{\frac{1}{k}+p}\right)^2\left(\frac{\frac{1}{k}-\frac{1}{p}}{\frac{1}{k}+\frac{1}{p}}\right)^2\bigr]\,.\\
%\end{multline}
$
\newline (Here we use dimensionless $x$ and $t$ coordinates : $x\to
mx$, $t\to mt$, thus the true time delay is obtained from the
dimensionless one presented below by dividing it by $m$). Analyzing
the $t\to\mp\infty$ limits of the solution and requiring that it
should correspond to the sum of two non interacting soliton antisoliton pairs
determines the $a_i$ $b_i$ $i=1,2$ parameters in terms of the initial ($t=t_0$)
soliton/antisoliton positions ($x_0^{is,\overline{s}}$) 
as well as the time delays: from the $t\to
-\infty$ limit it is found
\be\label{params}
 \eqalign{
{\scriptstyle a_1}&{\scriptstyle =-\gamma(x_0^{1{s}}+ut_0),}\qquad
{\scriptstyle a_2}{\scriptstyle =-\gamma(x_0^{1\overline{{s}}}-ut_0)+2\ln
u+\ln\left(\frac{\frac{1}{k}-p}{\frac{1}{k}+p}\right)^2
+\ln\left(\frac{\frac{1}{k}-\frac{1}{p}}{\frac{1}{k}+\frac{1}{p}}\right)^2\!,}\cr
{\scriptstyle b_1}&{\scriptstyle =-\tilde{\gamma}(x_0^{2{s}}+vt_0)+\ln\left(\frac{k-p}{k+p}\right)^2\!,}\qquad
{\scriptstyle b_2=-\tilde{\gamma}(x_0^{2\overline{{s}}}-vt_0)+2\ln
v+\ln\left(\frac{k-\frac{1}{p}}{k+\frac{1}{p}}\right)^2\!,
}}
\ee  
while the $t\to\infty$ limit yields the time delays of the two pairs
\be\label{idokes}
\eqalign{
\Delta t_1&=\frac{2\ln
u+\ln\left(\frac{\frac{1}{k}-p}{\frac{1}{k}+p}\right)^2+
\ln\left(\frac{k-p}{k+p}\right)^2}{\gamma u}\,, \label{e:4solik1}\cr
\Delta t_2&=\frac{2\ln
v+\ln\left(\frac{\frac{1}{k}-p}{\frac{1}{k}+p}\right)^2-
\ln\left(\frac{k-p}{k+p}\right)^2}{\tilde{\gamma} v}\,.
}
\ee
(The asymmetry in eq.(\ref{params}-\ref{idokes}) stems from assuming
$u>v$). These expressions for the time delay have a simple
interpretation: they give the sum of the time delays suffered in the
various collisions. Indeed the first terms on the right hand sides of
eq.(\ref{idokes}) give the time delays of the solitons from the
scattering on their own partners, while a simple Lorentz transformation
shows, that the second and third terms are nothing but the
contributions from the scattering on the two members of the other
pair.    

In the Neumann boundary problem the breather should be located at
$x=0$ and the soliton/antisoliton pair (representing the scattering
soliton) should also come together at the boundary. To accomplish this
the 4 soliton solution should be expressed in terms of the \lq
collision place' and \lq collision time' of each pair instead of the
initial positions. The collision place of each pair is trivially
$x^{*1}=(x_0^{1s}+x_0^{1\overline{s}})/2$, 
$x^{*2}=(x_0^{2s}+x_0^{2\overline{s}})/2$. Assuming that 
the slower moving  members of the inner
 pair collide first, 
the $t^{*1}$, $t^{*2}$ 
collision times can be obtained from the addition rule of the time
delays just shown, and the $a_i$, $b_i$
can be expressed more symmetrically using these four quantities:
\be\label{param2}
\eqalign{
{\scriptstyle a_1}&{\scriptstyle =-\gamma(x^{*1}+ut^{*1})+\ln
u+\ln\left(\frac{\frac1{k}-p}{\frac1{k}+p}\right)^2\!,}\qquad
{\scriptstyle a_2=-\gamma(x^{*1}-ut^{*1})+\ln
u+\ln\left(\frac{\frac1{k}-\frac1{p}}{\frac1{k}+\frac1{p}}\right)^2\!,}\cr
{\scriptstyle b_1}&{\scriptstyle =-\tilde{\gamma}(x^{*2}+vt^{*2})+\ln
v+\ln\left(\frac{k-p}{k+p}\right)^2\!,}\qquad 
{\scriptstyle b_2=-\tilde{\gamma}(x^{*2}-vt^{*2})+\ln
v+\ln\left(\frac{k-\frac1{p}}{k+\frac1{p}}\right)^2\!.
}}
\ee
Now the parameters of the
solution relevant for the Neumann problem are obtained as
follows: assuming we use the second pair to describe the breather 
we set $x^{*2}=0$ and continue $v$
to purely imaginary values $v=iw$ ($w$ real) and use eq.(\ref{param2})
to express the $b$ parameters; however the $a$ parameters are to be
obtained from eq.(\ref{params}) with
$x_0^{1s}=-x_0^{1\overline{s}}$. The reason behind this is that the
first two equations in (\ref{param2}) were obtained by assuming that
the soliton scatters on the individual members of the other pair, which is now
replaced by the breather. The time delay of the soliton is independent
of these parameters and 
 is obtained
from the first equation in (\ref{idokes}), which gives a real value in spite of $p$ being
a complex number:
\[
p=\sqrt{\frac{1+v}{1-v}}=\sqrt{\frac{1+iw}{1-iw}}=\frac{1+iw}{\sqrt{1+w^2}}=
{\rm e}^{i\arctan w}\,.
\]
Using this time delay in the integral in the semi-classical expression
(\ref{JaW}) gives
\be\label{JaW2}
 \frac{16}{\beta^2}\int\limits_0^{\tanh\theta}dv^\prime
\frac{\ln v^\prime}{1-v^{\prime
2}}+\frac{8}{\beta^2}\int\limits_0^k\frac{dy}{y}\left(\ln\left(\frac{y-p}{y+p}\right)^2+
\ln\left(\frac{y^{-1}-p}{y^{-1}+p}\right)^2\right)\,.
\ee
The first integral reproduces what is obtained above for ground state
boundary. In the second integral 
the $p$ parameter of the breather is obtained by matching the
classical and quantum expressions of its energy
\[
M\sin\left(\frac{\pi}{2\lambda}\right)=\frac{M}{\sqrt{1+w^2}}\,.
\]
Therefore in the semi-classical limit $p=i+\frac{\pi}{2\lambda}$; and
using it in the second integral shows that it is only an ${\cal
O}(\lambda^0)$ correction to the first one. Thus we verified the
matching between eq.(\ref{JaW}) and the limit of the exact amplitude
also in case of solitons reflecting on excited Neumann boundary.  

\subsection{Ground state boundary with general boundary conditions}

Finally we show that comparing the semi-classical limit of the exact
soliton/antisoliton reflection amplitude on the ground state boundary
with general boundary conditions and the semi-classical phase shift
obtained from eq.(\ref{JaW}) with the aid of the classical time delay
derived by SSW in \cite{SSW}, one can confirm the UV-IR relation
discussed in the previous section.

The most general reflection factor of the soliton
antisoliton multiplet \( |s,\bar{s}\rangle  \) on the ground state boundary,
satisfying the boundary versions of the Yang
Baxter, unitarity and crossing equations was found by Ghoshal and Zamolodchikov
\cite{GZ} as:
\begin{eqnarray*}
R(\eta ,\vartheta ,\theta ) & = & \left( \begin{array}{cc}
P^{+}(\eta ,\vartheta ,\theta ) & Q(\eta ,\vartheta ,\theta )\\
Q(\eta ,\vartheta ,\theta ) & P^{-}(\eta ,\vartheta ,\theta )
\end{array}\right) \nonumber \\
 & = & \left( \begin{array}{cc}
P_{0}^{+}(\eta ,\vartheta ,\theta ) & Q_{0}(\theta )\\
Q_{0}(\theta ) & P_{0}^{-}(\eta ,\vartheta ,\theta )
\end{array}\right) R_{0}(\theta )\frac{\sigma (\eta ,\theta )}{\cos (\eta )}\frac{\sigma (i\vartheta ,\theta )}{\cosh (\vartheta )}\, \, \, ,\nonumber \\
P_{0}^{\pm }(\eta ,\vartheta ,\theta ) & = & \cosh (\lambda \theta )\cos (\eta )\cosh (\vartheta )\pm i\sinh (\lambda \theta )\sin (\eta )\sinh (\vartheta )\nonumber \\
Q_{0}(\theta ) & = & i\sinh (\lambda \theta )\cosh (\lambda \theta )\,.\label{Rsas} 
\end{eqnarray*}
In \cite{SSW} useful integral representations are given for
\(R_0(\theta )\) and \(\sigma (x,\theta )\); for \(R_0(\theta )\) we
use this, while - by going back to the infinite product representation of
\cite{GZ} and \cite{genpap} - we replace
\[
\frac{\sigma (x,\theta )}{\cos x}=\frac{\Sigma (x,\theta )}{\cos
(x+i\lambda\theta )}
\]
with
 \[
\ln\Sigma
(x,\theta )=i\int\limits_0^\infty\frac{dy}{y}\frac{\sin(\frac{2\theta
y}{\pi})}{\sinh(y/\lambda)}\frac{\sinh(y-\frac{2x}{\pi\lambda}y)}{\cosh(y)}\,,
\]
as this gives a convergent integral in the entire range
\(0\le\eta\le\frac{\pi}{2}(\lambda +1)\). Expressing $\eta$ and
$\vartheta$ in terms of $c$ and $\vartheta_{\rm cl}$ as in section 3
and using the integral representations one obtains
\be\label{altkif}
R_0(\theta )\Sigma (\eta ,\theta )\Sigma(i\vartheta ,\theta)={\rm
e}^{i\hat{\delta}}{\rm e}^J,\qquad J=\int\limits_0^\infty \frac{dy}{y}
\frac{\sin\left(\frac{2y\theta}{\pi}\right)\sin\left(\frac{2y\vartheta_{\rm
cl}}{\pi}(\lambda^{-1}+1)\right)}{\sinh (y/\lambda)}\,.
\ee
In the semi-classical limit, neglecting the ${\cal O}(\lambda^0)$ terms
in the exponent
\[ {\rm e}^J\rightarrow \left\{ \begin{array}{c}
{\rm e}^{\lambda\vartheta_{\rm cl}}\qquad\theta >\vartheta_{\rm cl}\\
{\rm e}^{\lambda\theta}\qquad\theta <\vartheta_{\rm cl}
\end{array}\right. \, \, .\]
Therefore the three amplitudes, $P^\pm$ and $Q$, have rather different
semi-classical limits depending on whether the rapidity of the incident
particle is bigger or smaller than 
$\vartheta_{\rm cl}$:
\be\label{pqlim}
\eqalign{
\lim_{\lambda\to\infty}P^\pm=&{\rm e}^{\pm ic\frac{\pi}{2}\lambda}
{\rm e}^{ic\frac{\pi}{2}\lambda}{\rm e}^{i\hat{\delta}},\qquad
\lim_{\lambda\to\infty}Q=0,\qquad \theta <\vartheta_{\rm cl}\cr
\lim_{\lambda\to\infty}P^\pm=&0,
\quad\qquad\qquad
\lim_{\lambda\to\infty}Q={\rm e}^{ic\frac{\pi}{2}\lambda}{\rm
e}^{i\hat{\delta}},\qquad \theta >\vartheta_{\rm cl}\,.
}
\ee
This behaviour is consistent with the known facts, that classically, for Dirichlet
boundary conditions ($\vartheta_{\rm cl}=\infty $) solitons reflect as
solitons, while for Neumann boundary condition ($\vartheta_{\rm
cl}=0$) as antisolitons. Furthermore the classical solution found by
SSW \cite{SSW} shows the same critical behaviour as in
eq.(\ref{pqlim}), so that $\vartheta_{\rm cl}$ may be identified with
one of the parameters of that paper. To make the correspondence
complete one has to compute the semi-classical limit of $i\hat{\delta}$
as well. Using the aforementioned integral representations, after some
algebra, keeping only the leading terms, one finds:
\[
\eqalign{
\lim_{\lambda\to\infty}i\hat{\delta}&=-i\lambda\int\limits_0^\infty 
\frac{dy}{y^2}\sin\left(\frac{2\theta y}{\pi}\right)\left(\tanh (\frac{y}{2})+
\frac{\sinh ([c-1]y)}{\cosh y}+\tanh y -\tanh y\cos (\frac{2y\vartheta_{\rm
cl}}{\pi})\right)\cr &=-i(I_1+I_2+I_3+I_4)\,.
}
\]
All integrals $I_j$ are computed by realizing that $\frac{\partial
I_j}{\partial\theta}$ can be found in \cite{GR}. There is a subtlety
with $I_4$, as,
\[
 \frac{\partial
I_4}{\partial\theta}=\frac{\lambda}{\pi}\ln\left(\tanh\left[\frac{\theta+\vartheta_{\rm
cl}}{2}\right]\tanh\left[\frac{\vert\theta-\vartheta_{\rm
cl}\vert}{2}\right]\right),
\]
where $\vert\theta-\vartheta_{\rm cl}\vert $ is the modulus of $\theta-\vartheta_{\rm
cl}$. Therefore the $\theta<\vartheta_{\rm cl}$ and the
$\theta>\vartheta_{\rm cl}$ domains are separated by a logarithmic
singularity, and this matches nicely with eq.(\ref{pqlim}).  
Finally
\[
i\hat{\delta}=\frac{i\lambda}{\pi}\int\limits_{\theta_{\rm th}}^\theta dv\ln
\frac{\tanh^2 v\tanh^2 (v/2)}
{\tanh\left(\frac{1}{2}(v+i\frac{c\pi}{2})\right)
\tanh\left(\frac{1}{2}(v-i\frac{c\pi}{2})\right)
\tanh\left[\frac{v+\vartheta_{\rm
cl}}{2}\right]\tanh\left[\frac{\vert v-\vartheta_{\rm
cl}\vert}{2}\right]}\,,
\]
where $\theta_{\rm th}$ is $0$ in the $\theta<\vartheta_{\rm cl}$
domain, while it is $\vartheta_{\rm cl}$ in the $\theta>\vartheta_{\rm
cl}$ one. Now we are in a position to compare this to the integral of
the classical time delay derived in \cite{SSW}. SSW used two
parameters, $\zeta$ and $\eta_{SSW}$ (which we denote by $\hat{\chi}$
to avoid confusion) in
 that paper to describe the dependence of the time delay on the
Lagrangian parameters. These parameters are related to the Lagrangian
parameters of this paper by
\be\label{uuvir}
\eqalign{
2\cosh\zeta\cos\hat{\chi}&=-\frac{M_0\beta^2}{2m}\cos\alpha\,, \cr 
2\sinh\zeta\sin\hat{\chi}&=-\frac{M_0\beta^2}{2m}\sin\alpha\,.
}
\ee
Now making the shift $\hat{\chi}=\pi+\chi$ and the identifications 
\[
\chi
\to c\frac{\pi}{2}, \qquad\zeta\to\vartheta_{\rm cl},
\]  
converts on the one hand
the integral of the classical time delay in \cite{SSW} into
$\hat{\delta}$, while on the other it maps eq.(\ref{uuvir}) to our
previous UV-IR relation eq.(\ref{uvirr}-\ref{kvmcr}).\footnote{Note
that the $\zeta\to\vartheta_{\rm cl}$ identification is the same as
the one obtained from comparing the critical behaviour of the
classical solution \cite{SSW} and the limit of the quantum amplitude 
mentioned above.} Thus it is demonstrated that the UV-IR relation and
$M_{\rm crit}=\frac{4m}{\beta^2}$ in particular are also consistent
with the semi-classical soliton/antisoliton reflections.

\section{Conclusions}

In this paper two semi-classical issues of boundary sine-Gordon models
are investigated to get a better understanding of the relation between
the exact (algebraic) solution of the quantum theory and the classical
Lagrangian. 

First the semi-classical corrections to the energy difference of the two lowest energy 
static solutions were determined. In this procedure it turned out that
one has to renormalize also the boundary potential just in the same
way as the bulk one to obtain a finite result. 
Then we showed that comparing the main results of
the semi-classical quantisation - which include in addition to the
energy difference the semi-classical
bound states and the classical reflection factor of the sine-Gordon
field - and the semi-classical limit of the exact solution one can
obtain a relation between the Lagrangian and bootstrap parameters
provided we scale the bootstrap parameters in an appropriate way.  
After analytic continuation the
form of this relation coincides with what was found by Corrigan and
Taormina by semi-classically quantising the boundary breathers in
sinh-Gordon theory \cite{CT}. Since our computation is done in a
sector of sine-Gordon theory, which has no analogue in sinh-Gordon, 
this is an independent confirmation   of the results in \cite{CT}. We
also showed that in the semi-classical limit the UV-IR relation
obtained from describing the boundary sine-Gordon model as a bulk and
boundary  perturbed conformal field theory \cite{Zupb} coincides with
our result. 

Finally we analyzed the semi-classical soliton reflections building on
the ideas and results put forward by Saleur, Skorik and Warner
\cite{SSW}. As a consistency check we showed that the semi-classical
phase shift determined from the classical time delay and the number of
bound states agrees with the semi-classical limit of the exact
reflection amplitudes both for ground state and for the first excited
Neumann boundary. In the latter case we obtained the time delay from 
the analytic continuation of a
special two soliton - two antisoliton solution of the bulk theory,
that we constructed by the $\tau $ function method. Then we analyzed
the semi-classical limit of soliton/antisoliton reflections on ground state boundary with
general boundary conditions and confirmed the UV-IR relation
connecting the Lagrangian and bootstrap parameters. 

\subsubsection*{Acknowledgments}
   
We thank Z. Bajnok and G. Tak\'acs for the helpful discussions. 
This research was supported in part by the Hungarian Ministry
of Education under FKFP 0178/1999, and by the Hungarian
National Science Fund (OTKA) T029802/99, T34299/01.


\begin{thebibliography}{10}

\bibitem{SK}E.K. Sklyanin, {\sl Funct. Anal. Appl.} {\bf 21} (1987) 164.\\
            E.K. Sklyanin, {\sl J. Phys.} {\bf A21} (1988) 2375. 

\bibitem{GZ}S. Ghoshal and A. Zamolodchikov, \textsl{Int. J. Mod. Phys.} \textbf{A9}
(1994) 3841 (\texttt{hep-th/9306002}). 

\bibitem{FK}A. Fring and R. K\"oberle, \emph{Nucl. Phys.} \textbf{B421} (1994)
159 (\texttt{hep-th/9304141}).

\bibitem{gosh}S. Ghoshal, \emph{Int. J. Mod. Phys.} \textbf{A9} (1994)
4801 (\texttt{hep-th/9310188}). 

\bibitem{patr}P. ~Mattsson and P. ~Dorey, \emph{J. Phys}. \textbf{A33} (2000)
9065 (\texttt{hep-th/0008071}).\\
             P. ~Mattsson \lq\lq Integrable Quantum Field Theories, in
             the Bulk and with a Boundary'', Ph.D. thesis, hep-th/0111261.

\bibitem{genpap}Z. Bajnok, L. Palla, G. Tak\'acs and G.Zs. T\'oth, 
\emph{Nucl. Phys.} \textbf{B622} (2002) 548,  {\tt hep-th/0106070}.

\bibitem{Zupb}Al.B. Zamolodchikov, unpublished

\bibitem{uvir}Z. Bajnok, L. Palla and G. Tak\'acs, 
{\sl Nucl. Phys.} {\bf B622} (2002) 565, {\tt hep-th/0108157}.

\bibitem{CT}E. Corrigan and A. Taormina {\sl J. Phys.} {\bf A 33}
(2000) 8739 ({\tt hep-th/0008237}) 

\bibitem{SSW}H. Saleur, S. Skorik, N. P. Warner, {\it Nucl.~Phys.}
{\bf B441} (1995) 421, (\texttt{hep-th/9408004}). 

\bibitem{Raj}R. Rajaraman: Solitons and Instantons (North Holland 1982)

\bibitem{LL}L.D. Landau, E.M. Lifshitz: Theoretical Physics III:
Quantum Mechanics, Butterworth-Heinemann, 1997. 

\bibitem{dhn} R.F. ~Dashen, B. ~Hasslacher and A. ~Neveu, {\sl Phys. Rev.} 
{\bf D11} (1975) 3424.

\bibitem{Zoli} Z. Bajnok, G. B\"ohm and G. Tak\'acs in preparation 

\bibitem{Zamm}Al.B. Zamolodchikov, \emph{Int. J. Mod. Phys.}
\textbf{A10} (1995) 1125.

\bibitem{JW}R.~Jackiw, G.~Woo, {\it Phys.~Rev.} {\bf D12} (1975) 1643.
 
\bibitem{neupap}Z. Bajnok, L. Palla and G. Tak\'acs, {\sl Nucl. Phys.}
{\bf B614} (2001) 405, ({\tt hep-th/0106069}).

\bibitem{Abl}M. Ablowitz and H. Segur, Solitons and the inverse
scattering transform, Siam Stud. Appl. Math. 1981.

\bibitem{GR}I.S. Gradstein and I.M. Ryzhikh, Tables of Integrals,
Series and Products, V-th edition, Academic Press 1994.

\end{thebibliography}
\end{document}